# Exceptional unidirectional radiation modes

Shuowei An[1, 2], Tuo Liu[3, 4,*], Jie Zhu[5,*], Li Cheng[2,*]

[1] MOE Key Laboratory of Materials Physics and Chemistry under Extraordinary Conditions, Northwestern Polytechnical University, Xi'an, China

[2] Department of Mechanical Engineering, Hong Kong Polytechnic University, Hong Kong, China

[3] Institute of Acoustics, Chinese Academy of Sciences, Beijing 100190, China

[4] State Key Laboratory of Acoustics and Marine Information, Institute of Acoustics, Chinese Academy of Sciences, Beijing 100190, China

[5] Institute of Acoustics, School of Physics Science and Engineering, Tongji University, Shanghai 200092, China

[*] Corresponding authors. E-mail: liutuo@mail.ioa.ac.cn, jiezhu@tongji.edu.cn, and li.cheng@polyu.edu.hk.

## Abstract

Singularities such as exceptional points (EPs) and bound states in the continuum (BICs) fundamentally reshape the spectral response of non-Hermitian open systems, unlocking exotic physical phenomena and unprecedented ways of wave manipulation. Nevertheless, the radiation behaviors of modes evolving around these singularities lack a unified picture. Here we show that a mode originating from a BIC can radiate unidirectionally, exhibiting a quality factor that continuously varies in parameter space from a near-infinite value (near the BIC) to a finite lower bound as it coalesces with the other (leaky) mode at an EP. The crossing of the unidirectional radiation mode and the EP in parameter space results in not only merged poles (resonant EP) but also merged zeros (absorbing EP) of the scattering matrix, along with a counterintuitive reversal of radiation direction decoupled from the coalesced eigenmodes under near-field excitation. These phenomena are experimentally observed in an acoustic waveguide system featuring two side-branch resonant cavities with deliberate frequency detuning and asymmetric dissipation. Our findings reveal the intrinsic connection between eigenmode topology and radiation behavior in open systems, offering a viable route for tailoring scattering and radiation responses via non-Hermitian modulations.

## Introduction

Non-Hermitian systems are open systems that exchange energy with their surrounding environment via radiation and/or dissipation and can host exceptional points (EPs) appearing as branch-point singularities where eigenmodes coalesce with a defective eigenspace [1-6]. . EPs have attracted significant interest for the new insights they offer into the role of non-Hermiticity in mode interaction and evolution including mode braiding [7-9] and recovery [10-12], underpinning intriguing phenomena such as unidirectional invisibility [13, 14] and enhanced sensitivity [16-19], across diverse platforms ranging from quantum mechanics to optics and acoustics. Less attention, however, has been paid to the radiation of modes, especially at EPs. Existing studies either deal with the eigenspace per se with no accessible radiation channels included [7-12], or merely treat radiative losses as a source of non-Hermiticity for EP construction in open photonic crystal slabs [20], rather than manipulating the external radiated fields.

Of interest are other classes of spectral singularities in open systems, including bound states in the continuum (BICs) [21-25] and unidirectional radiation modes (URMs), i.e., unidirectional guided resonances (UGRs) in photonic crystal slabs [26-28]. Embedded within the continuous radiation spectrum, these two classes of states radiate uniquely, featuring radiation suppression across all channels or one of the two channels. They have been shown to arise from the mutual coupling of two modes [27, 29], e.g. Friedrich-Wintgen BICs induced by avoided energy-level crossings [30, 31], and UGRs originating from interband coupling [27]. Concerning the interplay between EPs and those peculiar radiative states,, previous studies have mainly focused on scattering EPs [28, 29, 32], non-Hermitian degeneracies of waveguide modes with opposite propagation or spin directions, or on resonant EPs constructed from two isolated BICs, wherein the radiation characteristics, however, are not necessarily correlated to the mode evolution [33].

In this letter, we consider a model as depicted in Fig. 1a and explore the formation and evolution of BICs, URMs and EPs within it. Two resonators characterized individually by onsite complex resonance frequencies, $\omega_a$ and $\omega_b$, are coupled (denoted by propagating-wave coupling $c$, expressed as $i\gamma_0 e^{i\phi}$) through their far-field radiation (denoted by radiative decay rate $\gamma_0$) into a shared bus waveguide. $\phi$ is the coupling phase modulated by the distance between the two resonators along the bus waveguide. Outgoing boundary conditions are imposed at the two ends of the waveguide, and the radiation to the left (right) port is denoted as $R_l$ ($R_r$).

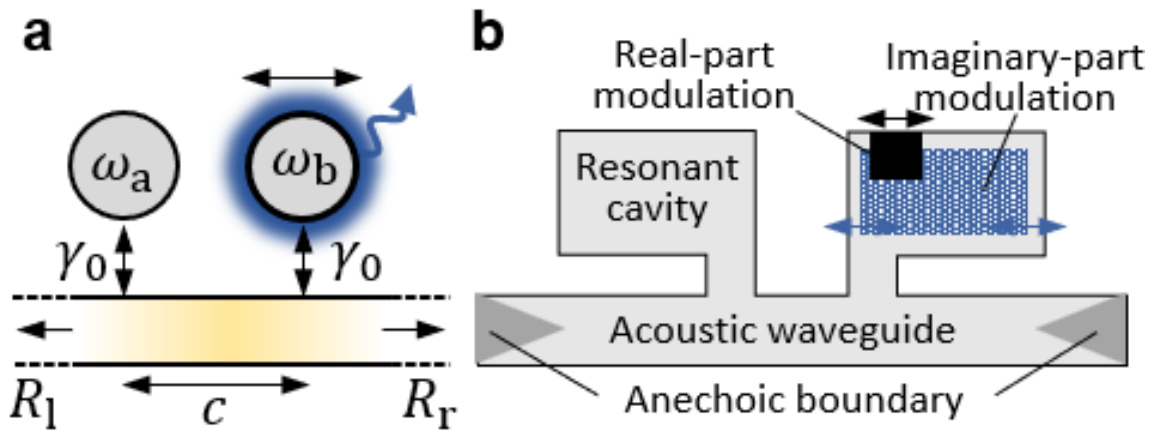

FIG. 1. Two-port, two-resonator-coupled non-Hermitian open system and its acoustical realization. (a) Model representation. (b) Schematic of the actual acoustic waveguide with two side-branch resonant cavities.

In addition to the coupling phase $\phi$ that has been extensively discussed in previous BIC studies [22, 34], we introduce a dimensionless complex onsite resonance frequency difference between the two resonators, $\delta\omega = (\omega_b - \omega_a)/\gamma_0$, where $\omega_a$ is fixed while the real and imaginary parts of $\omega_b$ are varied to produce desired frequency detuning and dissipation asymmetry (denoted by the horizontal arrows and blue shadow in Fig. 1(a), respectively). This difference breaks both the reflection and time-reversal symmetry of system, allowing us to examine the evolution of the eigenmodes and their radiation behaviors in the 3D parameter space spanned by the $\mathrm{Re}(\delta\omega)$, $\mathrm{Im}(\delta\omega)$ and $\phi$.

We implement the model experimentally using an acoustic waveguide with two side-branch resonant cavities (Fig. 1b). $\delta\omega$ is modulated within the complex domain: its real-part modulation is supplied by a movable acoustically rigid block inside the right cavity (which changes the resonance frequency of the cavity), while the imaginary-part one is achieved by a micro-perforated plate (MPP) with tunable effective area (which changes the dissipation strength), installed as the top face of the cavity (see Fig. 2b and experimental details in [35]). The required outgoing boundary conditions at the two ends are realized by sound-absorbing materials. Unlike photonic crystal slab systems, where exotic radiative states such as BICs or URMs are mostly obtained through structural or material modulation [26, 27, 36], our platform offers a much simpler configuration. It enables us to provide a rigorous theoretical proof for the existence of these states, the analytical derivation of their forming conditions and a comprehensive characterization of their evolution.

The radiation of the eigenmodes can be expressed as the superposition of the contributions from the two individual resonators,

$$\begin{pmatrix} R_\mathrm{l} \\ R_\mathrm{r} \end{pmatrix} = \mathbf{P\Psi} = \begin{bmatrix} \sqrt{\gamma_0} & \sqrt{\gamma_0}e^{-i\phi} \\ \sqrt{\gamma_0}e^{-i\phi} & \sqrt{\gamma_0} \end{bmatrix} \mathbf{\Psi} \tag{1}$$

where $\mathbf{R} = [R_\mathrm{l}, R_\mathrm{r}]^\mathrm{T}$ is the radiation vector and $\mathbf{\Psi}$ is the eigenvector of the system's Hamiltonian $\mathbf{H}$. $\mathbf{P}$ is a projection matrix that maps the eigenvector to the far-field radiation vector. This equation indicates that the radiation of an open system is essentially a projection of its eigenvectors and is therefore determined by both the eigenvectors themselves and how they project onto the radiation channels.

By letting $R_\mathrm{l} = 0$ or $R_\mathrm{r} = 0$, the condition of right radiation mode (RRM) or left radiation mode (LRM) is obtained in an analytical form as

$$\frac{1-e^{-2i\phi}}{i} - \delta\omega = 0 \ \text{(RRM), and} \ \frac{1-e^{2i\phi}}{ie^{2i\phi}} - \delta\omega = 0 \ \text{(LRM).} \tag{2}$$

Figure 2a presents the evolution of the two modes in the 3D parameter space $(\mathrm{Re}(\delta\omega), \mathrm{Im}(\delta\omega), \phi)$. Two dislocation lines of EPs, along with which both the eigenfrequencies, $\Omega_{1,2}$ and their associated eigenvectors, $\mathbf{\Psi}_{1,2}$, of the two-level system coalesce (i.e., $\Omega_1 = \Omega_2$, and $\mathbf{\Psi}_{1,2}$ become the same vector), are also included in Fig. 2a. It is seen that the two spiral dislocation lines corresponding to the LRM and RRM are situated within the two half spaces, $\mathrm{Im}(\delta\omega) > 0$ and $\mathrm{Im}(\delta\omega) < 0$, respectively. They intersect periodically with a period of $\pi$ to form BICs at $(0, 0, n\pi)$, $n \in \mathbb{Z}$. Meanwhile, the dislocation lines of EPs also follow a spiral trajectory given by $e^{i\phi}\delta\omega = \pm 2$, with a period of $2\pi$. Intriguingly, the dislocation lines of LRM and EPs intertwist in the 3D parameter space, and intersect at $(0, -2, n\pi + 0.5\pi)$ and $(0, 2, n\pi + 0.5\pi)$, as highlighted by the red and cyan dots, respectively. Such non-trivial intersections indicate that the two eigenmodes of the system coalesce and become URMs simultaneously at these points.

To further examine the two eigenmodes and their radiation, we select five cut-planes of $(\mathrm{Re}(\delta\omega), \mathrm{Im}(\delta\omega))$ at $\phi = 0, 0.25\pi,\ 0.5\pi,\ 0.75\pi$, and $\pi$, referred to as cut-planes I-V, in the parameter space (see the LRMs of the actual structures in [35]). As an intuitive illustration of the orientation between the two eigenmodes, here we define two normalized, real-valued vectors similar to those adopted for characterizing phase vector field in Ref. [27] as $[\mathrm{Re}(a_{1,2}), \mathrm{Im}(a_{1,2})]^{\mathrm{T}}/|[\mathrm{Re}(a_{1,2}), \mathrm{Im}(a_{1,2})]|$, where $a_{1,2}$ are extracted from the two eigenvectors in the form $\boldsymbol{\Psi}_{1,2} = [a_{1,2}, 1]^{\mathrm{T}}$. These two normalized vectors are mapped into the two arrows in Fig. 2b, serving as a measure of the linear dependence of two eigenmodes. The two arrows point in the same direction at EPs but in opposite directions at BICs, corresponding to parallelism and orthogonality of the two eigenvectors, respectively. Figure 2c shows the real (top panel) and imaginary (bottom panel) parts of the two eigenfrequencies at the five cut-planes, in which the eigenmode merging forms two EPs. The surfaces are color-coded by the radiation contrast, $\log|R_l| - \log|R_r|$, which reaches its minimum for the RRM and its maximum for the LRM, respectively. Figures 2b and 2c at cut-plane III further display the coincidence between the EPs and the URMs (see Ref. [35] for the topological index of the radiation phase singularity).

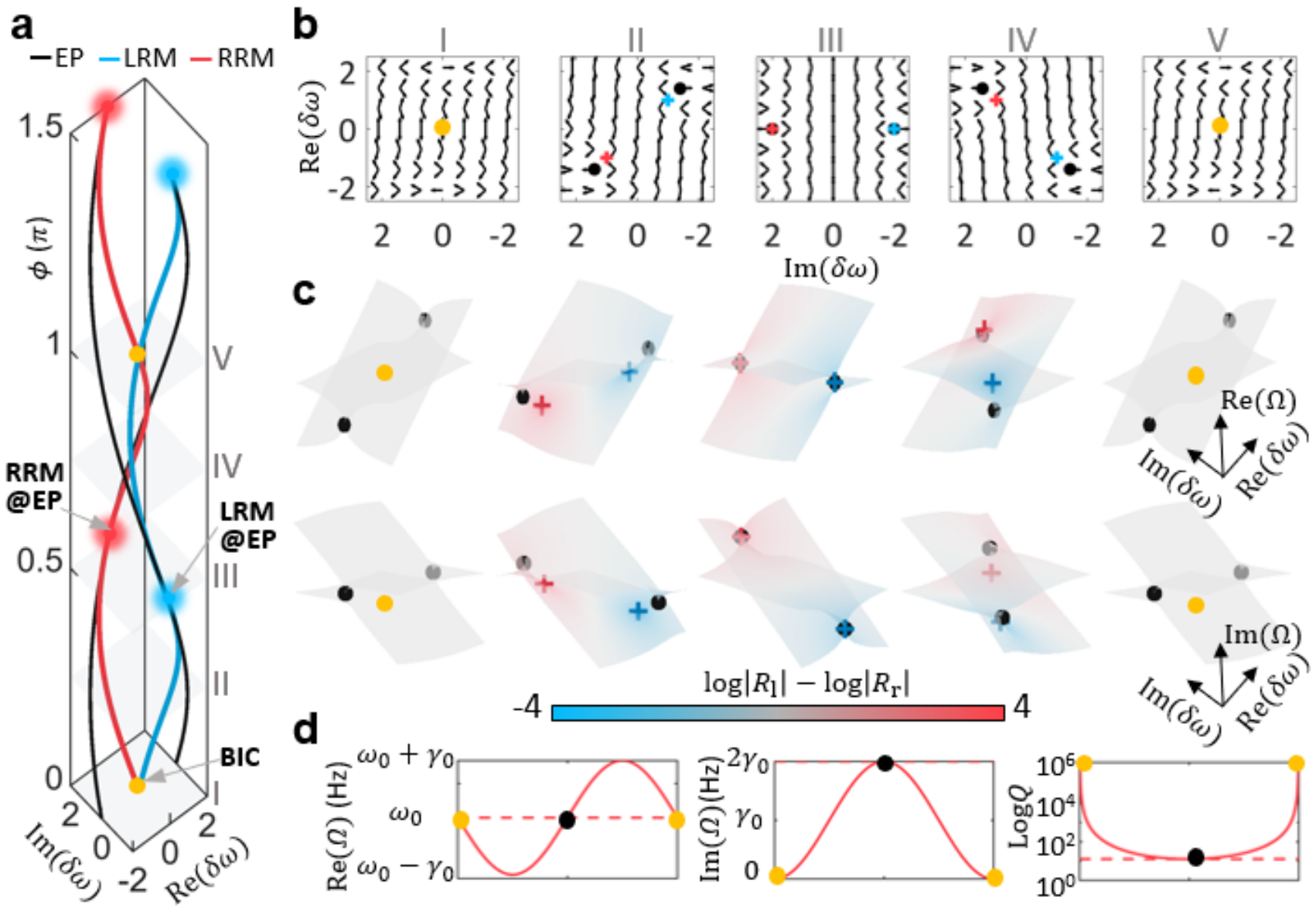

FIG. 2. Mode evolution in a 3D parameter space spanned by the frequency differences

between the two complex onsite eigenfrequencies and the coupling phase. (a) Dislocation lines of exceptional points (EPs, black lines), right radiation mode (RRM, cyan lines) and left radiation mode (LRM, red lines) evolving in a 3D parametric space. The intersections of RRM and LRM are BICs (orange dots). (b) Distributions of representative normalized vectors, at five cut-planes I, II, III, IV and V. The arrows depict the two representative normalized vectors derived from the two eigenvectors. (c) Real (top panel) and imaginary (bottom panel) parts of the eigenfrequencies at the five selected cut-planes. The color scale represents the radiation contrast between the left and right ports. Black dots, cyan cross and red cross denote EPs, RRM and LRM, respectively. (d) Eigenfrequencies of the two eigenmodes including LRM and the accompanying leaky modes, and their $Q$ factors along the dislocation lines as $\phi$ ranges from 0 to $\pi$.

In contrast to the orthogonality between a BIC and its accompanying leaky mode [cut-planes I and V], the URMs and their accompanying leaky modes are no longer orthogonal with mutual coupling [cut-planes III-IV], which in turn gives rise to varied $Q$ factors of the URM across the parameter space. We take the LRM as an example. The coupling is the weakest around the BICs, where radiation leakage is suppressed, endowing the LRM with a nearly infinite $Q$ factor there, i.e., $\lim_{\phi\to 0^+} Q_{\mathrm{LRM}} = \frac{\mathrm{Re}(\omega)}{2\mathrm{Im}(\omega)} = \infty$, as evidenced by Fig. 2(d). The coupling strength gradually increases and reaches its maximum when the two eigenmodes coalesce at EPs at $\phi = 0.5\pi$, resulting in the drop of the $Q$ factor to a finite value, $\omega_0/(4\gamma_0)$.

Mathematically, the asymmetric radiation of the system originates from the rank behavior of projection matrix, $\mathbf{P}$. When $\phi \neq n\pi$, $\mathbf{P}$ has full rank ($\mathrm{rank}(\mathbf{P}) = 2$), enabling a complete mapping from the eigenspace $\mathbf{\Psi}$ [Fig. 2b, cut-planes II-IV] to the radiation vector space $\mathbf{R}$ via Eq. (1). This results in a variation in the radiation contrasts and their corresponding extremes [Fig. 2c, cut-planes II-IV)]. Conversely, when $\phi = n\pi$, $\mathbf{P}$ becomes rank-deficient ($\mathrm{rank}(\mathbf{P}) = 1$), collapsing the original two-dimensional eigenspace $\mathbf{\Psi}$ into a one-dimensional radiation space, $\mathbf{R}$. Consequently, the radiation contrast becomes constant across the entire Riemann surfaces [Fig. 2c, cut-planes I and V], taking a value of zero in our system.

The scattering and resonant properties of a system are fully encoded in the eigenvalue poles and zeros of its scattering matrix distributed across the complex frequency plane [25]. Poles inherently correspond to the eigenfrequencies of $\mathbf{H}$ under outgoing boundary conditions, whereas zeros represent vanishing scattered fields upon a specific incident wave front. Here, we characterize the BICs and the URMs through a complex-frequency analysis of the scattering of the system under far-field excitation, with parameters adopted according to the dislocation line of LRM in Fig. 3a. The LRM rather than the RRM is chosen because the former requires only asymmetric loss, which is more feasible than the asymmetric gain required by the latter. Five cut-planes at 0, $0.25\pi$, $0.5\pi$, $0.75\pi$, and $\pi$ are selected for this analysis.

As shown in Figs. 3a and 3b, when $\phi = 0$ (cut-plane I), there exists a pair of conjugate pole and zero ($P_1$ and $Z_1$ in Fig. 3b) as well as a pair of invisible pole and zero canceled out with each other at the real axis ($P_2$ and $Z_2$), suggesting a leaky mode and a BIC. Once deviates from $\phi = 0$, $P_1$ is fixed while $Z_1$ varies, and the previously invisible $P_2$ and $Z_2$ associated with the BIC depart from each other and now represent the LRM, as exemplified by cut-plane II at $\phi = 0.25\pi$. In this case, the pole-zero conjugation is no longer guaranteed due to the broken time-reversal symmetry caused by the added intrinsic loss of the right resonator ($\mathrm{Im}(\delta\omega) > 0$) [25]. These poles and zeros vary with a period of $\pi$. Remarkably, when $\phi = 0.5\pi$, i.e., the LRM at EP, not only do the two poles coalesce, but so do the two zeros. The former, named a resonant EP [37], results from the coalescence of the eigenmodes of $\mathbf{H}$, as discussed above. The latter is known as an absorbing EP and usually considered as a sign of a coherent perfect absorbing (CPA) EP when located exactly at the real frequency axis [37, 38]. However, in our system, the coalesced eigenvector of the scattering matrix $\mathbf{S}$ is $[0, 1]^{\mathrm{T}}$, signifying one-sided perfect absorption from the right port rather than conventional CPA, which requires coherent incoming waves injected from both two ports, (i.e., incident wave fronts $[1, 1]^{\mathrm{T}}$ or $[1, -1]^{\mathrm{T}}$). This represents an extreme case of CPA EP with the selective suppression of radiation channels [35].

The corresponding real-frequency reflection spectra ($r_{\mathrm{r}}$ and $r_{\mathrm{l}}$) obtained from CMT with $\phi$ continuously varying from 0 to $\pi$ are further presented in Fig. 3c. Three representative spectra at $\phi = 0$, $0.25\pi$, and $0.5\pi$ are provided in the top panels of Fig. 3d, in which the transmission spectra ($t$) are also included. In the CMT calculations, $\omega_0$ and $\gamma_0$ are set as 1532 Hz and 16 Hz, respectively, extracted from the actual cavity sample in both simulations and experiments.

When $\phi = 0$, the system hosts a BIC and a leaky mode. The BIC is fully decoupled from the far field and therefore invisible in the spectrum (Fig. 3d, left panel). The remaining leaky mode solely determines the reflection spectrum. In this case, the spectrum behaves like that of an open one-level system with uniform reflections for both directions. Once $\phi$ deviates from 0, the two eigenmodes become non-orthogonal: an LRM and a leaky mode. The LRM acts as a "unidirectional BIC", which is excited solely from the left-hand side with a finite linewidth. The leaky mode, on the other hand, leads to reflection peaks for both incidences, as shown in the middle panel of Fig. 3d (see more results in Ref. [35]). As $\phi$ increases to $0.5\pi$, the two eigenmodes coalesce as a single LRM. Consequently, the system is excited only for left incidence but muted for right incidence, resulting in a coalesced reflection peak in $r_{\mathrm{l}}$ but a complete absence of reflection in $r_{\mathrm{r}}$ across the whole spectrum (Fig. 3d, right panel). During this continuous variation in $\phi$, the two eigenmodes interact more strongly, from being decoupled to coalesced, and the linewidth of LRM in $r_{\mathrm{l}}$ changes from infinitely narrow (infinite $Q$ factor) to the widest with $Q = \omega_a/(4\gamma_0)$ (Fig. 3c).

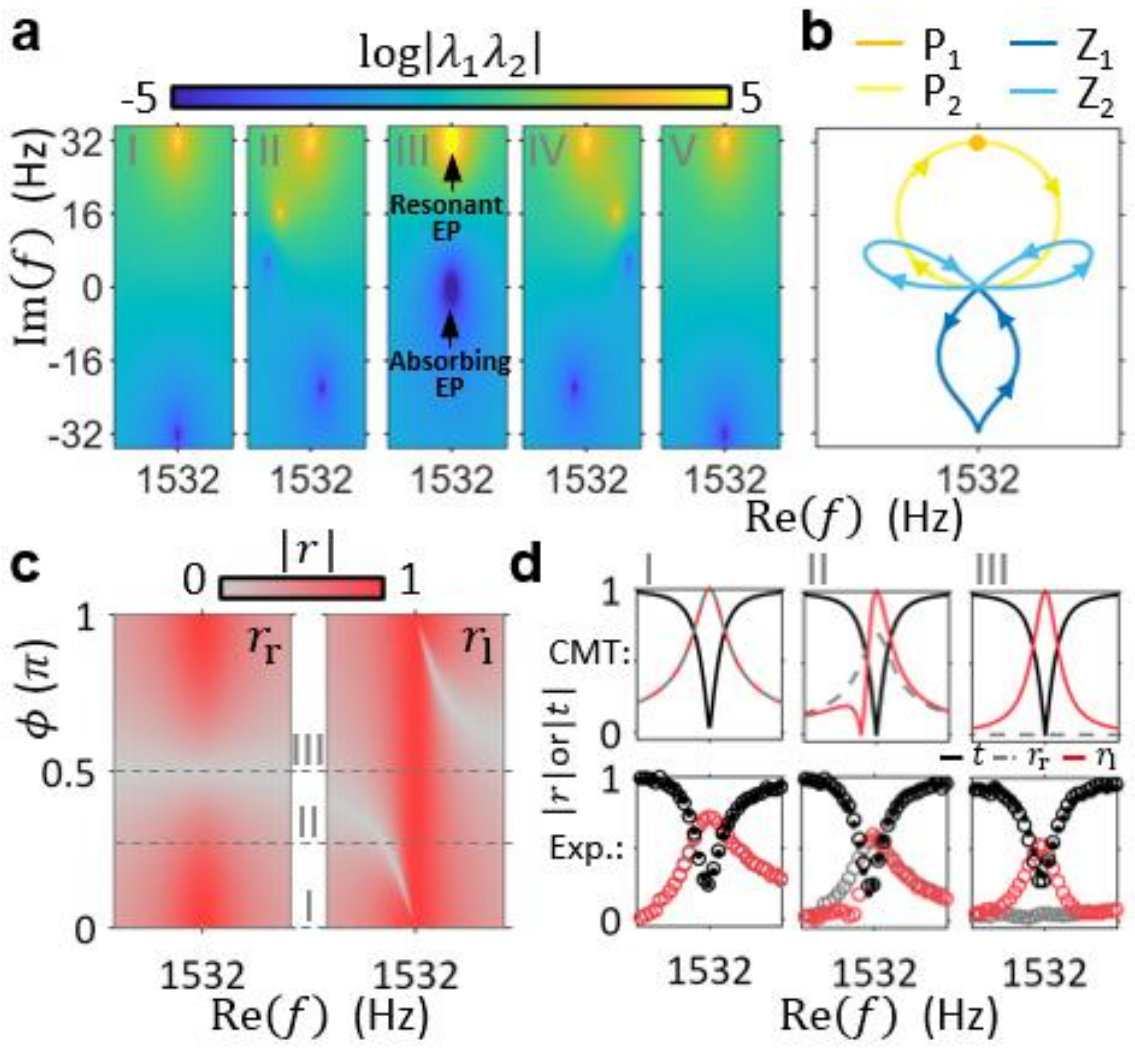

FIG. 3. Scattering characteristics of the LRMs. (a) Complex-frequency plane analysis of the LRM at five representative cut-planes corresponding to I-V in Fig. 3a. (b) Trajectories of two scattering zeros ($Z_1$, $Z_2$) and poles ($P_1$, $P_2$) as the system parameters are swept along the LRM dislocation line Fig. 3a. Arrows denote the evolution directions as $\phi$ varies from 0 to $\pi$. (c) Reflection coefficients for right (top panel) and left (top panel) incidences versus real frequency and $\phi$. (d) Three selected spectra I, II and III corresponding $\phi = 0$, $0.25\pi$ and $0.5\pi$ (marked by black dashed lines in (c)). The upper and lower panels show the scattering coefficients obtained from the CMT and experiments, respectively.

The results from the experiments are also presented in the lower panels of Fig. 3d for comparisons and validations (see additional simulation results in Ref. [35]). These results agree well with the CMT results, except for the lower peak amplitudes of $r_{\mathrm{l}}$, and $r_{\mathrm{r}}$ attributed to the inevitable thermal-viscous loss inside the cavity [35].

A more direct approach for eigenmode characterization is to place near-field sources and probe the system's radiation response. To excite the LRM exclusively without activating its accompanying leaky mode, we employ a pair of coherent sources. According to the eigenvectors of the LRM, the two sources placed at two resonators have equal amplitude but a phase difference, i.e., $\boldsymbol{S}_{\mathbf{near}} = \left[e^{i(\phi-\pi)}, 1\right]^{\mathrm{T}}$.

The radiation responses of the LRMs at different cut-planes with $\phi$ from 0 to $\pi$

obtained from CMT are shown in Figs. 4(a)-4(b). As expected, at $\phi = 0$ (cut-plane I), a BIC with vanishing linewidth radiates to neither the left nor the right ports. Once $\phi$ deviates from 0, the BIC becomes an LRM that radiates exclusively to the left port, with no radiation to the right. These LRMs are distinguished by their $Q$ factors, i.e., the closer the LRM is to the BIC, the higher its $Q$ factor (see section 3 in [35]). Their $Q$ factor has no theoretical upper limit but a lower one when the LRM coincides with an EP at $\phi = 0.5\pi$. Measured radiation spectra at $\phi = 0$, $0.25\pi$ and $0.5\pi$ agree well with the CMT results (bottom panel of Fig. 4b, cut-planes I-III, and section 7 in [35]). The small radiation peaks observed in the measured spectra (black circles in Fig. 4b), possibly arising from weak radiation losses induced by imperfect sample fabrication [35], remain much lower than those of the LRM.

What is more interesting is the radiation response of the system under single source near-field excitation at EP, where the two eigenmodes coalesce as one LRM. Without loss of generality, we place the source at either the left or the right resonator to excite the system.

The results from the CMT and experiments (Fig. 4c) consistently show that the excited radiation fields cannot be fully described by the coalesced or defective eigenspace. The radiation can even occur in the opposite direction to the eigenmode. More specifically, when the source is positioned at the right cavity, the excited field radiates both leftwards and rightwards, a behavior clearly beyond the eigenspace that contains only the coalesced LRM. When the source is positioned at left cavity, the field radiates abnormally rightward, behaving like an RRM instead of its eigenmode (LRM). We term this phenomenon as direction-reversal radiation, which is analogous to the chirality-reversal radiation observed in a passive parity-time-symmetric ring cavity [39]. This decoupled radiation behavior can be explained mathematically by a Jordan vector that complements the defective eigenspace of $\mathbf{H}$ at the EP [39, 40].

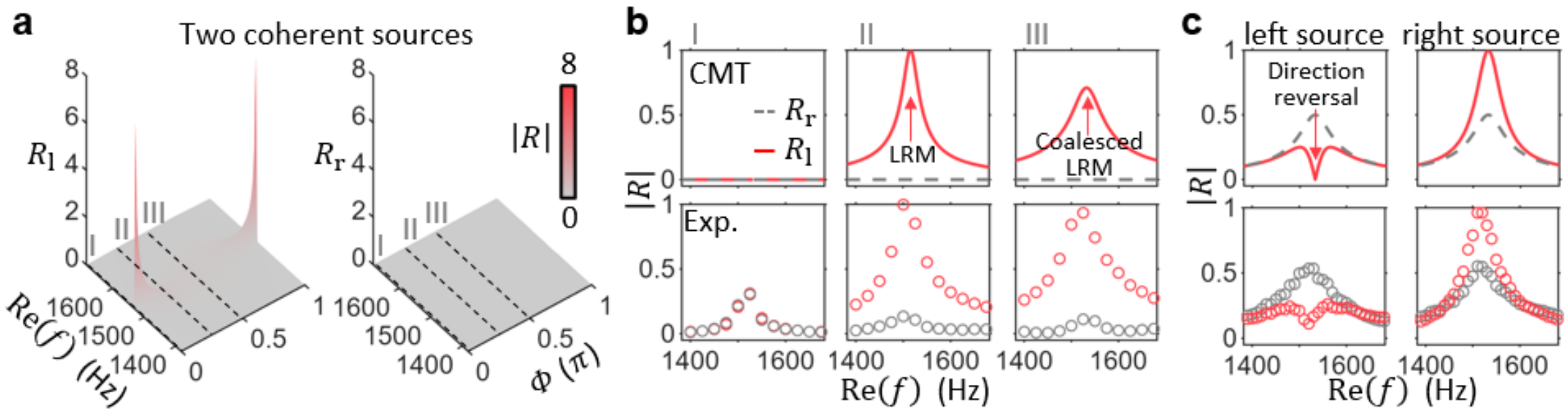

FIG. 4. Radiation characteristics of LRMs excited by two coherent sources and one single source. (a) Left and right radiation spectra of LRMs as $\phi$ varies from 0 to $\pi$. The color scale represents their amplitudes. (b) Three selected spectra I, II and III corresponding to $\phi = 0,\ 0.25\pi$ and $0.5\pi$. The top and bottom panels show the radiation spectra obtained from the CMT and experiments, respectively. (c) Radiation spectra obtained from CMT (top panel) and experiments (bottom panel) with a single source positioned at the left (left panel) and right (right panel) cavities.

To conclude, we have systematically demonstrated the emergence of URMs in a two-level, two-port open acoustic system and their evolution as the system approaches EPs. A particular coincidence of EP and URM is found, which gives birth to both resonant EP and absorbing EP, as well as a direction-reversal radiation phenomenon. Our work provides a new perspective for understanding unconventional scattering and radiation responses ‘of non-Hermitian systems. This approach can be extended to other wave systems, such as elastic, optical and magnonic systems, and may find promising applications in wave amplification and absorption.

**Acknowledgements**

We thank Yuanze Li from Hong Kong Polytechnic University for assistance with the acoustic experiments, and Dr. Lei Xiao from Guangzhou University for discussions on the graphical presentation of this paper. This work was supported by the National Natural Science Foundation of China (Grant No. 12604677) and the Basic and Frontier Exploration Project Independently Deployed by Institute of Acoustics, Chinese Academy of Sciences (Grant No. JCQY202403).

## End Matter

At EPs, the coalescence of eigenmodes can give rise to higher-order poles in the system response, leading to spectral features beyond the conventional first-order Lorentzian line shape. We investigate these features in the reflection and radiation spectra of our system when the left radiation mode (LRM) coincides with an EP at $(0, 2, \frac{\pi}{2})$ in parameter space.

The mode coalescence at an EP leads to a square Lorentzian spectrum of left reflection,

$$r_{\mathrm{l}} = -4i\gamma_0^2/(\omega - \omega_a - 2i\gamma_0)^2, \tag{A1}$$

compared with the first-order Lorentzian spectrum of the two-port system coupled with one single resonator, $r = i\gamma_0/(\omega - \omega_a - i\gamma_0)$.

The radiation spectra of the system at EP also exhibit squared Lorentzian line shapes, as exemplified by the case of a source in the left cavity, which take the form

$$\begin{cases} R_{\mathrm{r}} = \frac{2\gamma_0^{1.5} + i(\omega-\omega_a)\gamma_0^{0.5}}{(\omega-\omega_a-2i\gamma_0)^2} \\ R_{\mathrm{l}} = \frac{4\gamma_0^{1.5} + i(\omega-\omega_a)\gamma_0^{0.5}}{(\omega-\omega_a-2i\gamma_0)^2}. \end{cases} \tag{A2}$$

They are further compared with the spectrum for a two-port system with a single side-branch resonator, $R = \frac{\gamma_0^{0.5}}{\gamma_0 + i(\omega_a - \omega)}$, which peaks at $\omega = \omega_a$ with amplitude $\gamma_0^{-0.5}$. Remarkably, the additional intrinsic loss in the right cavity necessary for EP construction does not suppress radiation; the peak magnitude of $R_{\mathrm{l}}$ remains identical to that of the one-level system. Moreover, the mode coalescence yields a broadened higher-order Lorentzian line shape, indicating that tailoring multiple coupled QNMs can enable broadband radiation enhancement [41, 42].